\documentstyle[11pt]{article}
\def\Journal#1#2#3#4{{#1} {\bf #2}, #3 (#4)}

\def\NPB{{\em Nucl. Phys.} B}
\def\PLB{{\em Phys. Lett.} B}
\def\PRL{\em Phys. Rev. Lett.}
\def\PRD{{\em Phys. Rev.} D}

\def\MPL{{\em Mod. Phys. Lett.} A}

\def\l{\ell}

\def\q{\hat{q}}
\def\s{\hat{s}}

\def\mc{\hat{m}_c}
\def\muq{\hat{m}_u}
\def\mq{\hat{m}_q}

\title{Determination of the CKM matrix by semileptonic 
	rare B-decays \footnote{Talk presented at APCTP 
	Workshop : Pacific Particle Physics Phenomenology
	in Seoul, Korea (October 31 $\sim$ November 2, 1997).}}

\author{L. T. Handoko \thanks{On leave from P3FT-LIPI, Indonesia. 
        E-mail : handoko@theo.phys.sci.hiroshima-u.ac.jp.} \\
	Department of Physics, Hiroshima University\\
	1-3-1 Kagamiyama, Higashi Hiroshima - 739, Japan}
\date{}

\begin{document}

\setlength{\baselineskip}{24pt}

\maketitle
\begin{picture}(0,0)
       \put(325,160){HUPD-9725}
       \put(325,145){December 1997}
\end{picture}
\vspace{-24pt}

\thispagestyle{empty}
\begin{abstract}
The possibility to test the unitarity triangle of CKM 
matrix is examined by using the semileptonic rare $B$
decays, $B \rightarrow X_q \, \l^+ \, \l^-$ with $q = d, s$ 
and $\l$'s are leptons. The discussion is emphasized on
the CP asymmetry in the framework of standard and beyond the
standard models. For the beyond standard model, a minimal 
extension of the standard model containing an additional 
isosinglet charge (-1/3) charge quark, which leads to a 
deviation from CKM unitarity, is considered. As the results,
in the standard model the CP asymmetry is visible only for 
$B \rightarrow X_d \, \l^+ \, \l^-$, while in the latter 
model it is sizeable even for $B \rightarrow X_s \, \l^+ \,
\l^-$.
\end{abstract}

\section{Introduction}
\label{sec:intro}

After the measurement of the radiative 
$B \rightarrow X_s \, \gamma$ decay, the next target for 
the experiments is other higher order decays, e.g. 
$B \rightarrow X_s \, \l^+ \, \l^-$. The CLEO Collaboration's 
study indicates to be able to measure few events number of this mode in near
future \cite{bsll}.

Theoretically, the decays are very atractive and should be
good probes to test the standard model as well as open a
window to the theory beyond it. In general the rare decays,
that means the decays which occur in one-loop level or more,
are very sensitive to the new contributions. Here according
to the unitarity triangle of CKM matrix, the new physics 
can be divided into two types, 1) the new physics which
conserve the unitarity, and 2) the new physics which violate it.
In this proceeding, I consider only the standard model and 
the latter type of new physics that can be realized by
introducing an additional isosinglet charge (-1/3) charge quark.

As the measurement that must be examined in the experiments,
the interest is focused on the CP asymmetry. Remark that, in 
the decay which is governed by single operator like
$B \rightarrow X_s \, \gamma$, the CP asymmetry is tiny as 
indicated, for example, in some previous papers \cite{lth0}. 
In another word, $B \rightarrow X_s \, \gamma$ decay is not 
useful to extract the CP violation. Then one needs to consider 
more promising modes. The best candidate is the semileptonic 
$B \rightarrow X_q \, \l^+ \, \l^-$ decays with $q = d, s$.

The reason is very clear, that is 
$B \rightarrow X_q \, \l^+ \, \l^-$ decay contains minimally
three operators in its effective hamiltonian, that is 
\begin{eqnarray}
        {\cal H}_{\rm eff} & = & \frac{G_F \, \alpha}{\sqrt{2} \, \pi} \, 
                V_{tq}^\ast \, V_{tb} \, \left\{ 
                {C_9}^{\rm eff} \, 
                        \left[ \bar{q} \, \gamma_\mu \, L \, b \right] \, 
                        \left[ \bar{\l} \, \gamma^\mu \, \l \right]
                + {C_{10}}^{\rm eff} \,  \left[ \bar{q} \, \gamma_\mu \, L \, b		\right] \, 
                        \left[ \bar{\l} \, \gamma^\mu \, \gamma_5 \, \l \right]
                \right. \nonumber \\
        & &     \; \; \; \; \; \; \; \; \; \; \; \; \; \; \;
		\; \;  \; \; \; \; \; \; \left.
               - 2 \, {C_7}^{\rm eff} \, 
                        \left[ \bar{q} \, i \, \sigma_{\mu \nu} \, 
                        \frac{\q^\nu}{\s} \left( R + \mq \, L \right) \, b
                        \right] 
                        \left[ \bar{\l} \, \gamma^\mu \, \l \right]
                \right\} \; .
                \label{eq:heff}
\end{eqnarray}
Here, $C^{\rm eff}_i$ denotes the Wilson coefficient for
each operator, $q^\mu$ is four-momentum of the dilepton 
and $s = q^2$. Therefore, the CP violation will be occured 
if there is different phases in, at least, one of the Wilson
coefficients. 

\section{Conserved CKM unitarity case}
\label{sec:ckmuni}

Now, I am not going to give a specific model of new physics 
that conserves the unitarity of CKM matrix, but discuss only
the case in the framework of standard model. 

In the standard model, some different phases are possibly 
appearing in the long-distance contributions of the decay.
The reason is because $m_b$ is heavy enough to generate on-shell states of
uponium and/or charmonium, i.e. $u^i \bar{u}^i$ with $u^i = u, c$.
These $u^i \bar{u}^i$ loops are generated from  the 
$b \rightarrow q \, u^i \, \bar{u}^i$ processes that is 
governed by the following operators, 
\begin{eqnarray}
        {\cal O}_1 & = & 
                \left( \bar{q}_\alpha \, \gamma_\mu \, L \, b_\alpha 
                        \right) \, 
                \sum_{i=1,2} \, 
                \left( \bar{u}_\beta^i \, \gamma^\mu \, L \, u_\beta^i 
                        \right) \; , 
                \label{eq:o1} \\
        {\cal O}_2 & = & 
                \left( \bar{q}_\alpha \, \gamma_\mu \, L \, b_\beta
                        \right) \, 
                \sum_{i=1,2} \, 
                \left( \bar{u}_\beta^i \, \gamma^\mu \, L \, u_\alpha^i 
                        \right) \; , 
                \label{eq:o2}
\end{eqnarray}
after doing Fierz transformation. Here, $u^1 = u$, $u^2 = c$ and the lower
suffixes denote the color.
For $q = s$, $u\bar{u}$ loop contribution can be ignored 
\cite{bqll},
while for $q = d$ the situation is quite different, since \cite{lth1}
\begin{eqnarray}
        q = s : 
        \left| \frac{V_{u^is}^\ast \, V_{u^ib}}{V_{ts}^\ast \, V_{tb}} \right| 
                & \sim & \left\{
                \begin{array}{lcl}
                        O(\lambda^2)    & , & u^i = u \\ 
                        O(1)            & , & u^i = c
                \end{array}
                \right. 
        \nonumber \\
        q = d : 
        \left| \frac{V_{u^id}^\ast \, V_{u^ib}}{V_{td}^\ast \, V_{tb}} \right| 
                & \sim & \left\{ 
                \begin{array}{lcl}
                        O(1)            & , & u^i = u \\
                        O(1)            & , & u^i = c
                \end{array}
                \right. 
        \nonumber
\end{eqnarray} 
if we normalize the amplitude with $V_{tq}^\ast \, V_{tb}$ as usual.
$\lambda$ is a parameter in Wolfenstein parametrization of 
CKM matrix \cite{wolfenstein} and the world average is
$\lambda \sim 0.22$ \cite{pdg}.

\begin{figure}[t]
        \unitlength 0.7mm
        \begin{center}
        \begin{picture}(60,40)
                \put(5,5){\vector(0,1){50}}
                \put(5,5){\vector(1,0){75}}
                \thicklines
                \put(65,5.1){\line(-1,0){60}}
                \put(5,5){\line(1,2){20}}
                \put(25,45){\line(1,-1){40}}
                \multiput(25,5)(0,5){8}{\line(0,1){1.8}}
                \multiput(5,45)(5,0){4}{\line(1,0){1.8}}
                \put(5,2){\makebox(0,0){$(0,0)$}}
                \put(65,2){\makebox(0,0){$(1,0)$}}
                \put(25,47){\makebox(0,0){$(\rho,\eta)$}}
                \put(25,38){\makebox(0,0){$\alpha$}}
                \put(57,8){\makebox(0,0){$\beta$}}
                \put(10,8){\makebox(0,0){$\gamma$}}
                \put(25,2){\makebox(0,0){$(\rho,0)$}}
                \put(-2,45){\makebox(0,0){$(0,\eta)$}}
                \put(12,27){\makebox(0,0){$r$}}
                \put(50,27){\makebox(0,0){$x$}}
        \end{picture}
        \caption{The CKM unitarity triangle on the $\rho-\eta$ plane.}
        \label{fig:ckm}
        \end{center}
\end{figure}
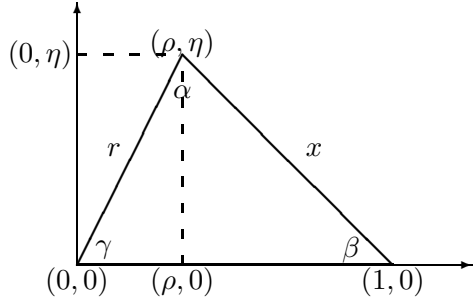

Further, if the triangle is parametrized as Fig. \ref{fig:ckm},
then one finds 
\begin{equation}
        \frac{V_{uq}^\ast \, V_{ub}}{V_{tq}^\ast \, V_{tb}} \sim 
        \left\{
        \begin{array}{lcl}
                \frac{\displaystyle 
                r \left( e^{-i \gamma} - r \right)}{
                \displaystyle 1 + r^2 - 2 \, r \, \cos \gamma}  & , & q = d \\
                & & \\
                \lambda^2 \, r \, e^{-i \gamma} & , & q = s \; \; \; \; 
                \mbox{\tiny $(\lambda^2 \sim 5\%)$} 
        \end{array}
        \right. 
\end{equation}
From the result, it is obvious that for $q = d$ the uponium 
($u\bar{u}$) is large enough to induce the CP violation in
the decay. On the other hand, for $q = s$ the CP violation 
will be suppressed by a factor of $\lambda^2$.
These long-distance effects lead the Wilson coefficients 
to be \cite{lth1}, 
\begin{eqnarray}
        {C_7}^{\rm eff} & = & {C_7}^{\rm NLO} \, ,
	\label{eq:c7sm} \\
        {C_9}^{\rm eff} & = & {C_9}^{\rm NLO} + 
                \frac{V_{uq}^\ast \, V_{ub}}{V_{tq}^\ast \, V_{tb}} \, ,
                {C_9}^{\rm CP}
	\label{eq:c9sm} \\
        {C_{10}}^{\rm eff} & = & C_{10} \, ,
	\label{eq:c10sm}
\end{eqnarray}
where ${C_7}^{\rm eff}$ and ${C_9}^{\rm eff}$ have been 
calculated up to next-to-leading order \cite{qcdc9} \cite{qcdc7}, 
while ${C_{10}}^{\rm eff}$ 
obtains no correction at all. The ${C_9}^{\rm CP}$ is the 
CP violation factor and given as, 
\begin{eqnarray}
        {C_9}^{\rm CP} & = &
                \left( 3 \, C_1 + C_2 + 3 \, C_3
                        + C_4 + 3 \, C_5 + C_6 \right)
		\nonumber \\
        & & \times \left[ \displaystyle\underbrace{
                g(\mc, \s) - g(\muq, \s)}_{\mbox{\tiny continuum}} 
		- \frac{16 \, \pi^2}{9} 
                \displaystyle\underbrace{\left(
                \sum_{V=\psi,\cdots} F_V(\s) - 
                \sum_{V=\rho,\omega} F_V(\s) \right)}_{
                \mbox{\tiny resonances}} \right] \, .
\end{eqnarray}
The expressions for the continuum and resonances contribution 
can be seen in reference \cite{lth1}.

\section{Violated CKM unitarity case}
\label{sec:vckmuni}

For the case when the CKM unitarity is violated, I adopt 
a definite extension of the standard model containing an 
additional isosinglet charge (-1/3) charge quark \cite{vlq}. 

In this model, the gauge sector are modified like below, 
\begin{eqnarray}
        {\cal L}_{W^{\pm}} & = & \frac{g}{\sqrt{2}} \, 
                \underline{V_{i \alpha}} \bar{u}_i \, \gamma^\mu \, L \,
                d_\alpha \, W_{\mu}^+ + {\rm h.c.} \, , \\
        {\cal L}_Z & = & \frac{g}{2 \cos \theta_W} \, 
                \bar{d}_{\alpha} \, \gamma^{\mu} \left[
                \left( \frac{2}{3} \sin^2 \theta_W \, \delta_{\alpha\beta} -
                \underline{z_{\alpha \beta}} \right) L 
                + \frac{2}{3} \sin^2 \theta_W \, \delta_{\alpha\beta} \, 
                R \right] d_\beta \, Z_{\mu} \, ,
\end{eqnarray}
where $z_{\alpha \beta} \equiv \sum_{i=1}^{3} {U^d}_{\alpha i}
{U^d}_{i \beta}^{\ast} 
= \delta_{\alpha\beta} - {U^d}_{\alpha 4} {U^d}_{4 \beta}^{\ast}$
and the new $3\times4$ CKM matrix satisfies the relation 
$ \sum V_{iq}^\ast \, V_{ib} = z_{qb}$ that shows 
the violated unitarity for $z_{qb} \neq 0 $.

As long as one considers only the tree-level contributions 
of new physics, in this case is the $Z-$exchange tree-level
diagram, the Wilson coefficients would read \cite{lth2},
\begin{eqnarray}
	{C_7}^{\rm eff} & \rightarrow & {C_7}^{\rm eff} \\
	{C_9}^{\rm eff} & \rightarrow & {C_9}^{\rm eff} 
		+ \left( 1 + \frac{\alpha_s(\mu)}{\pi} \omega(\s) \right) 
		\frac{\pi}{\alpha} \, \frac{z_{qb}}{V_{tq}^\ast \, V_{tb}} 
                \left( 4 \, \sin^2 \theta_W - 1 \right) \\
	{C_{10}}^{\rm eff} & \rightarrow & {C_{10}}^{\rm eff} +
		\frac{\pi}{\alpha} \, \frac{z_{qb}}{V_{tq}^\ast \, V_{tb}}
\end{eqnarray}
The problem is then how to determine the size of the new mixings 
${z_{qb}}/{V_{tq}^\ast \, V_{tb}}$. Fortunately, in principle the upper
bounds for the mixings can be extracted from the 
experimentally known $B \rightarrow X_s \, \gamma$ decay for 
$q = s$ and from $B_d^0-\bar{B}_d^0$ mixing for $q = d$ \cite{lth2}.

\section{Summary}
\label{sec:summary}

By including the long-distance contributions in the 
$B \rightarrow X_q \, \l^+ \, \l^-$ decays, there will 
be additional phases in the amplitude. These should 
induce the CP asymmetry that is defined as,
\begin{equation}
                \bar{\cal A}_{\rm CP} = \frac{\displaystyle
                {{\rm d}{\cal B}}/{{\rm d}\s}
                - 
                {{\rm d}\bar{\cal B}}/{{\rm d}\s}
                }{\displaystyle 
                {{\rm d}{\cal B}}/{{\rm d}\s}
                + 
                {{\rm d}\bar{\cal B}}/{{\rm d}\s}
                }       
                = \frac{\displaystyle
                -2 \, {{\rm d}{\cal A}_{\rm CP}}/{{\rm d}\s}
                }{
                {{\rm d}{\cal B}}/{{\rm d}\s} + 
                2 \, {{\rm d}{\cal A}_{\rm CP}}/{{\rm d}\s}},
\end{equation}
with ${\cal B} = {\rm Br}(B \rightarrow X_q \, \l^+ \, \l^-)$ 
and $\bar{\cal B}$ is its complex conjugate. 

Lastly, as the results in both cases I found : 
\begin{enumerate}
\item In the standard model where the CKM unitarity is conserved, 
	$\bar{\cal A}_{\rm CP}(B \rightarrow X_s \, \l^+ \, \l^-) 
	\sim {\rm few} \times {10}^{-3}$ and 
	$\bar{\cal A}_{\rm CP}(B \rightarrow X_d \, \l^+ \, \l^-) 
	\sim 5\%$.
\item In the beyond standard model where the CKM unitarity is 
	violated, 
	$\bar{\cal A}_{\rm CP}(B \rightarrow X_s \, \l^+ \, \l^-) 
	\sim 2\%$ and 
	$\bar{\cal A}_{\rm CP}(B \rightarrow X_d \, \l^+ \, \l^-) 
	\sim 5\%$, by including only the $Z-$exchange tree diagram.
\end{enumerate}
Therefore, one can say that once the CP asymmetry in 
$B \rightarrow X_s \, \l^+ \, \l^-$ is detected, it 
will be a strong proof of the new physics beyond the standard
model.

\section*{Acknowledgments}

The author would like to thank Organizing Committee of the Workshop
for their warm hospitality during the stay in Seoul, Korea.
A part of this work is supported by Ministry of Education 
and Culture of Japan under Monbusho Fellowship Program.


\begin{thebibliography}{99}
	\bibitem{bsll} Y. J. Kwon (CLEO Collaboration), 
	in {\em this Workshop Proceeding} (World Scientific, 1998).
        \bibitem{lth0}
                G. Bhattacharyya, G. C. Branco and D. Choudhury,  
                \Journal{\PLB}{336}{487}{1994}.
                [Err. \Journal{\PLB}{340}{266}{1994}]; \\
                L. T. Handoko and T. Morozumi, 
                \Journal{\MPL}{10}{309}{1995}.
	\bibitem{lth1} L. T. Handoko to appear in 
		{\it Phys. Rev.} {\bf D} (1998)
		(hep-ph/9707222).
        \bibitem{bqll}
		B. Grinstein, M. J. Savage and M. B. Wise, 
		\Journal{\NPB}{319}{271}{1989}; \\
		R. Grigjanis, P. J. O'Donnel, M. Sutherland and H. Navelet, 
		\Journal{\PLB}{223}{239}{1989}.
	\bibitem{qcdc9}
		M. Je$\dot{\rm z}$abek and J. H. K$\ddot{\rm u}$hn, 
		\Journal{\NPB}{320}{20}{1989}; \\
		M. Misiak, 
		\Journal{\NPB}{393}{23}{1993}
		[Err. {\it ibid.} {\bf B439}, 461 (1995)]; \\ 
		A. J. Buras and M. M$\ddot{\rm u}$nz,
		\Journal{\PRD}{52}{186}{1995}.
	\bibitem{qcdc7}
		A. J. Buras, A. Kwiatkowski and N. Pott, 
		(hep-ph/9710336).
	\bibitem{wolfenstein}
		L. Wolfenstein, 
		\Journal{\PRL}{51}{1945}{1983}.
        \bibitem{pdg}
                Particle Data Group, 
                \Journal{\PRD}{54}{1}{1996}.
        \bibitem{vlq}
                G. C. Branco and L. Lavoura, 
                \Journal{\NPB}{278}{738}{1986}; \\
                L. Lavoura and J. P. Silva, 
                \Journal{\PRD}{47}{2046}{1993}; \\
                V. Barger, M. S. Berger and R. J. N. Philips,
                \Journal{\PRD}{52}{1663}{1995}.
	\bibitem{lth2} L. T. Handoko, 
		submitted to {\it Phys. Lett.} {\bf B} (1998)
		(hep-ph/9708447).
\end{thebibliography}
\end{document}